\documentclass[12pt,a4paper]{amsart}

\usepackage[latin1]{inputenc}
\usepackage{amsmath,amssymb,amsthm,bm}
\usepackage{graphicx}
\usepackage{subfigure}

\numberwithin{equation}{section}

\newtheorem{proposition}{Proposition}

\theoremstyle{definition}

\newtheorem{example}{Example}
\newtheorem{remark}{Remark}
\newtheorem{theorem}{Theorem}
\newtheorem{lemma}{Lemma}

\newcommand{\hi}{\mathcal{H}} 
\newcommand{\lh}{\mathcal{L(H)}} 
\newcommand{\trh}{\mathcal{T(H)}} 
\newcommand{\sh}{{\mathcal{T(H)}^+_1}} 
\newcommand{\ket}[1]{|#1\rangle} 
\newcommand{\kb}[2]{|#1\rangle\langle#2|} 
\newcommand{\tr}{\textrm{tr\,}} 
\newcommand{\fii}{\varphi}
\newcommand{\SR}{\mathcal{S}}
\newcommand{\cal }{\mathcal}

\newcommand{\R}{\mathbb R}
\newcommand{\C}{\mathbb C}

\newcommand{\N}{\mathbb N}
\newcommand{\Z}{\mathbb Z}
\renewcommand{\P}{{\mathbb P}^2}
\renewcommand{\S}{{\mathbb S}^1}
\newcommand{\Ra}{\mathsf R}
\renewcommand{\H}{\mathsf H}

\newcommand{\bor}[1]{\mathcal{B}({#1})}

\newcommand{\E}{\mathsf{E}} 
\newcommand{\Q}{\mathsf{Q}} 
\newcommand{\Eh}{\mathsf{E}_{\rm ht}} 
 
\def\<{\langle}
\def\>{\rangle}

\begin{document}

\title[Quantum tomography and generalized Markov kernels]{Quantum tomography, phase space observables, and generalized Markov kernels}
\author{Juha-Pekka Pellonp\"a\"a}
\address{Juha-Pekka Pellonp\"a\"a,
Department of Physics and Astronomy, University of Turku,
FIN-20014 Turku, Finland}
\email{juhpello@utu.fi}
\begin{abstract}
We construct a generalized Markov kernel which transforms 
the observable associated with the homodyne tomography into 
a covariant phase space observable with a regular kernel state.
Illustrative examples are given in the cases of a 'Schr\"odinger cat' kernel state and the Cahill-Glauber $s$-parametrized distributions. Also we consider an example of a kernel state when the generalized Markov kernel cannot be constructed.
\newline

\end{abstract}
\maketitle

\section{Introduction}
Quantum homodyne tomography and 8-port homodyne detection are important tools in quantum optics for reconstruction 
of a quantum state of light (see, \cite{Le} and references therein).
In both measurements, there are associated observables, normalized positive operator measures (POMs), which
give the measurement outcome statistics of the input state.
In the 8-port (or double) detection, the corresponding observable is a covariant phase space observable $\E_K$ generated
by a kernel state $K$ (see, e.g.\ \cite{KiLa}).
Recently, Albini et al.\ \cite{AlDeTo} found a POM $\Eh$ associated with
the quantum homodyne tomography (see also \cite{KiLaPe}).

The aim of this paper is to find a connection between the two POMs $\Eh$ and $\E_K$ when $K$ is regular enough.
We will show that $\E_K$ is connected to $\Eh$ via a generalized Markov kernel. A drawback is that the generalized Markov kernel is not a positive function, so that, it cannot be interpreted as a transition probability from $\Eh$ to $\E_K$.
However, $\E_K$ is a postprocessing of $\Eh$ and thus subordinate to the tomographic measurement $\Eh$.

This work is a direct continuation of our earlier work \cite{LaPe} where we considered only the Husimi $Q$-function case $K=\kb00$. We also showed that there is no hope to represent $\Eh$ as a postprocessing of $\E_{\kb00}$ at least by using generalized Markov kernels.

The structure of this article is the following:
In section 2, we introduce the basic notations, definitions and some well-known results. 
We also recall the structures of the POMs $\Eh$ and $\E_K$ associated with the tomographic and 8-port homodyne measurements.
We will use the Radon and Hilbert transformations which are probably familiar to readers so that their properties needed in this article are collected in the Appendix.

Section 3 contains the main results of this paper and some illustrative examples. 
First we note that the tomographic data is somehow 'sharper' than the 8-port data for the given input state $T$.
Then we construct the generalized Markov kernels for POMs $\E_K$ (with suitable kernel states $K$) and then expand them to Hermite and Maclaurin series. Both series are used in the examples. 
The first example shows the structure of the kernel associated to the phase space observable generated by a 'Schr\"odinger cat' state. In the second example, kernels are constructed for Cahill-Glauber $s$-parametrized distributions (or more precisely, for operator measures associated to them). We will see that
although the Husimi $Q$-distribution has a very simple kernel, the Wigner function has no kernel at all.
Finally, we see that there does not necessarily exist a generalized Markov kernel for an arbitrary kernel state $K$.
\section{Notations and definitions}

For any measure space $(\Omega,\mu)$, where $\mu$ is a positive measure on $\Omega$, we let 
$L^1(\Omega)$ (resp.\ $L^2(\Omega)$) denote the space of equivalence classes of integrable (resp.\ square integrable) functions $\Omega\to\C$.
In the case of the Hilbert space $L^2(\Omega)$, we let $\<\,\cdot\,|\,\cdot\,\>_{L^2(\Omega)}$ denote its innerproduct which is assumed to be linear with respect to the second argument.
When $\Omega=\R^n$, $n=1,2$, the measure $\mu$ is always the $n$-dimensional Lebesgue measure.
We let $\SR(\R^n)$ be the Schwartz space of rapidly decreasing smooth complex functions on $\R^n$. We identify $\SR(\R^n)$ with a subset of $L^1(\R^n)\cap L^2(\R^n)$.

Let  $\hi$ be a complex Hilbert space with an innerproduct $\<\,\cdot\,|\,\cdot\,\>$ and
an orthonormal (number) basis $\{\ket n\,|\,n\in\N\}$ where $\N:=\{0,1,2,...\}$. 
Let $\lh$ (resp.\ $\trh$) be the set of bounded operators (resp.\ trace-class operators) on $\hi$. The set of states (density matrices), that is, the positive operators $T\in\trh$ of trace 1, is denoted by $\sh$.
For any $B\in\lh$, we denote $B_{mn}:=\<m|B|n\>$, $m,\,n\in\N$, so that $B=\sum_{m,n=0}^\infty B_{mn}\kb m n$ where the double series converges with respect to the weak operator topology.

Define the
lowering, raising, and number operators 
$$
A:=\sum_{n=0}^\infty\sqrt{n+1}\kb n{n+1},\hspace{1cm} 
A^*:=\sum_{n=0}^\infty\sqrt{n+1}\kb{n+1}n,\hspace{1cm} 
N:=A^*A=\sum_{n=0}^\infty n\kb n n ,
$$
respectively. 
Physically the Hilbert space $\hi$ and the above operators are associated to a single-mode optical field.
We will, {\it without explicit indication,} use the
coordinate representation, in which $\hi$ is represented as $L^2(\R)$ via the
unitary map $ |n\rangle\mapsto h_n $, where $h_n\in\SR(\R)$ is the $n$th Hermite function,
$$
h_n(x) := \frac{1}{\sqrt{2^nn!\sqrt{\pi}}}H_n(x)e^{-\frac 12 x^2}
=\frac{(-1)^n}{\sqrt{2^nn!\sqrt{\pi}}} e^{\frac12x^2}{d^n e^{-x^2}\over dx^n}
$$
and $H_n$ is the $n$th Hermite polynomial.\footnote{Hermite polynomials are given
by the recursion relation $H_0(x)=1$, $H_1(x)=2x$, and $H_{n+1}(x)= 2xH_n(x)-2nH_{n-1}(x)$,
or by the Rodrigue's formula $H_n(x)=(-1)^n e^{x^2}d^n e^{-x^2}/dx^n$.}
Recall that it is customary to denote (formally) $\psi(x)=\<x|\psi\>$ and $\int_\R|x\>\<x|dx=I$ (the identity operator), that is,
$\<\psi|\fii\>=\int_\R\<\psi|x\>\<x|\fii\>dx=\int_\R\overline{\psi(x)}\fii(x)dx$ for all $\psi,\,\fii\in\hi\cong L^2(\R)$.

Define\footnote{$\overline{B}$ means the closure of a linear operator $B$ and $\overline z$ is a complex conjugate of $z\in\C$.} the position operator $Q:=\frac{1}{\sqrt{2}}\overline{(A^*+A)}$ and the momentum operator
$P:=\frac{i}{\sqrt{2}}\overline{(A^*-A)}$, which, in the coordinate
representation are the usual multiplication and differentiation operators, respectively:
$(Q\psi)(x) = x\psi(x)$ and $(P\psi)(x)= -i\frac{d\psi}{dx}(x)$.
For any $\theta\in[0,2\pi)$,
$$
Q_\theta:=(\cos\theta) Q+(\sin\theta) P
$$ 
is the (self-adjoint) {\it rotated quadrature operator} and 
$\Q_\theta:\,\bor\R\to \lh $ is the spectral measure of $Q_\theta$; 
we denote the Borel $\sigma$-algebra of any topological space $\Omega$ by $\cal B(\Omega)$.
It is easy to see that $\Q_\theta(R)=e^{i\theta N}\Q_0(R)e^{-i\theta N}$ for all $R\in\bor\R$ and $\theta\in\R$. 

Define the Weyl operator (or the displacement operator of the complex plane)\footnote{Recall that Weyl operators are associated to a unitary representation of the Heisenberg group $\mathbb H$, or to a projective representation of $\C\cong\R^2$.}  
$D(z)=e^{\overline{z A^*-\overline z A}}$ for which 
$$
D(z)^*=D(z)^{-1}=D(-z)\hspace{1cm}
\text{ and } \hspace{1cm}
e^{i\theta N}D(z)e^{-i\theta N}=D\big(ze^{i\theta}\big)
$$ 
where $z\in\C$ and $\theta\in\R$. In addition, for all $z_1,\,z_2\in\C$,
 \begin{equation}\label{1}
 D(z_1+z_2)=e^{-i\,{\rm Im}(z_1\overline{z_2})}D(z_1)D(z_2).
 \end{equation}
 Let $q,\,p\in\R$ and $z=(q+i p)/\sqrt2$. Then
 $$
 D(q,p):=D(z)=e^{ipQ-iqP}=e^{-iqp/2}e^{ipQ}e^{-iqP}=e^{iqp/2}e^{-iqP}e^{ipQ},
 $$
 that is, for all $\psi\in\hi\cong L^2(\R)$, $(e^{ipQ}\psi)(x)=e^{ipx}\psi(x)$, 
 $(e^{iqP}\psi)(x)=\psi(x+q)$, and
 $$
 \big(D(q,p)\psi\big)(x)=e^{-iqp/2}e^{ipx}\psi(x-q).
 $$
 Using the polar coordinate parametrization $z=\rho e^{i\theta}$, $\rho=|z|=\sqrt{\frac{1}{2}(q^2+p^2)}$, $\theta=\arg z=\arctan(p/q)$, that is, $q=\sqrt{2}\rho\cos\theta$, $p=\sqrt{2}\rho\sin\theta$, one gets
 \begin{equation}\label{4}
e^{i\sqrt 2 \rho Q_\theta}=e^{i(\sqrt 2 \rho\cos\theta) Q+i(\sqrt 2 \rho\sin\theta) P}=e^{iqQ+ipP}=D(-p,q).
 \end{equation}
Moreover, the two-dimensional Lebesgue measure is
$$
d^2z=\rho d\rho d\theta=\frac{1}{2}d\rho^2d\theta=\frac{1}{2}dqdp.
$$


\subsection{Covariant phase space observables and Wigner functions}

Let $K$ be
a trace-class operator on $\hi$. Define an operator measure $\E_K:\,\cal B(\C)\to \lh $ as follows: for all $Z\in\cal B(\C)$,
$$
\E_K(Z):=\frac1\pi\int_ZD(z)KD(z)^*d^2z=\frac1{2\pi}\int_ZD(q,p)KD(q,p)^*dqdp.
$$
Since $\E_K(\C)=(\tr K)I$ we see that $\E_K$ is normalized if and only if $\tr K=1$. Moreover, $\E_K$ is positive if and only if $K$ is a positive operator.
 If $K\in\sh$ we say that the normalized positive operator measure (POM) $\E_K$ is a 
{\it covariant phase space observable} generated by the kernel $K$.
In principle, any covariant phase space observable can be measured by using double homodyne detection 
(see, e.g.\ \cite{KiLa} and references therein).

The parity operator is
$$
{\Pi}:=R(\pi)=\sum_{n=0}^\infty (-1)^n|n\rangle\langle n|={\Pi}^+-{\Pi}^-
$$
where ${\Pi}^+:=\sum_{k=0}^\infty|2k\rangle\langle 2k|$ and ${\Pi}^-:=\sum_{k=0}^\infty|2k+1\rangle\langle 2k+1|$
are projections onto the closed subspaces $\hi^+:={\Pi}^+\hi$ and $\hi^-:={\Pi}^-\hi$ (i.e., $\hi=\hi^+\oplus\hi^-$).
Elements of $\hi^+$ (resp.\ $\hi^-$) are even (resp.\ odd) functions.
For any $\psi\in\hi\cong L^2(\R)$ we can write
$$
({\Pi}^{\pm}\psi)(x)=\frac{1}{2}\big[\psi(x)\pm\psi(-x)\big]
$$
so that $({\Pi}\psi)(x)=\psi(-x)$ or, formally, ${\Pi}=\int_\R|x\>\<-x|dx$.

The Wigner function $W^T:\,\C\to\C$ of a trace-class operator $T$ is
\begin{eqnarray}\label{Wigner}
&&\\
\nonumber 
&&W^T(z')\equiv W^T(q',p'):=\frac1\pi\tr[TD(z'){\Pi}D(z')^*]=\frac1\pi\tr[T D(2 z'){\Pi}] \\
&&=
\frac{1}{4\pi^2}\int_{\R^2}\tr[TD(-p,q)]e^{-iq'q-ip'p}dqdp
=
\frac{1}{4\pi^2}\int_0^{2\pi}\int_0^\infty\tr[Te^{i\sqrt 2 \rho Q_\theta}]e^{-i\sqrt{2}\rho(q'\cos\theta+p'\sin\theta)}d\rho^2d\theta
\nonumber
\end{eqnarray}
where $z'=(q'+ip')/\sqrt2\in\C$
(see, equation \eqref{4}).
Thus, the probability measures $R\mapsto\tr[T\Q_\theta(R)]$, $\theta\in[0,2\pi)$, fully determine the Wigner function $W^T$ of any $T\in\sh$.
If $W^T$ is integrable\footnote{This need not hold even for all pure states $T=\kb\eta\eta$ (see example 3).} then
$$
\int_{\R^2} W^T(q,p)dqdp=2\int_\C W^T(z)d^2z=\tr[T].
$$
Recall that the Wigner function is always square integrable and 
\begin{equation}\label{2}
{\rm tr}[TK]=2\pi\int_{\R^2}W^T(q,p)W^K(q,p)dqdp=4\pi \int_\C W^T(z)W^K(z)d^2 z.
\end{equation}
In addition,
$$
W^T(q,p)=\frac{1}{\pi}\int_\R \tilde{T}(q-x,q+x)e^{2ipx}dx
$$
where $(x,y)\mapsto \tilde T(x,y)$ is the integral kernel\footnote{That is, for all $\psi\in L^2(\R)$,
$(T\psi)(x)=\int_\R \tilde T(x,y)\psi(y)dy$ for $dx$-almost all $x\in\R$. More precisely, $\tilde T$ is an equivalence class of functions which satisfy this condition.}
 of a trace-class operator $T$ considered as an integral operator on $L^2(\R)$; usually one denotes formally $\tilde T(x,y)\equiv\<x|T|y\>$.
For future use, we define a linear space
$$
\cal T^S(\hi):=\{T\in\trh\,|\,\tilde T\in\cal S(\R^2)\}.
$$ 
Note that, the mapping $\cal T^S(\hi)\ni T\mapsto W^T\in\cal S(\R^2)$ is bijective (see, e.g.\ prop.\ 4 of \cite{AlDeTo}).
 
For any $K$ and $T\in\trh$ we get
$$
\tr{[T\E_K(Z)]}=\frac1\pi\int_Z\tr{[TD(z_1)KD(z_1)^*]}d^2z_1=\frac1\pi\int_Zp^T_K(z_1)d^2z_1
$$
where, by equations \eqref{2} and \eqref{1}, the density
\begin{eqnarray*}
p^T_K(z_1)&:=&\tr{[TD(z_1)KD(z_1)^*]}=4\pi \int_\C W^{TD(z_1)}(z_2)W^{KD(z_1)^*}(z_2)d^2 z_2 \\
&=&
4\pi \int_\C W^T(z_1/2+z_2)W^K(-z_1/2+z_2)d^2 z_2
=4\pi \int_\C W^T(z)W^K(z-z_1)d^2 z
\end{eqnarray*}
and $p^T_K(z_1)=p^K_T(-z_1)$. Note that $p^T_K$ is a well-defined continuous function even if $K$ is only a bounded operator. Thus, from equation \eqref{Wigner}, one has
$$
W^T=\frac{1}{2\pi}p^T_{2\Pi}.
$$


\subsection{The POM associated to homodyne tomography}
Let $\Omega:=[0,\pi)\times\R$ so that $\Omega$ gives a parametrization for the projective space $\P$ (see, Appendix).
Define the normalized positive operator measure (POM) $\widetilde\Eh:\,\bor\Omega\to \lh $
associated to {\bf h}omodyne {\bf t}omography \cite{AlDeTo} as
$$
\widetilde\Eh(\Theta\times R):=\frac{1}{\pi}\int_\Theta \Q_\theta(R)d\theta
=\frac{1}{\pi}\int_\Theta e^{i\theta N}\Q_0(R)e^{-i\theta N}d\theta,\hspace{1cm}\Theta\in\bor{[0,\pi)},\;R\in\bor\R.
$$
Obviously, one can define similarly a POM associated to the double covering $\S\times\R$ of $\P$ as
$$
\Eh(\Theta\times R):=\frac{1}{2\pi}\int_\Theta \Q_\theta(R)d\theta=
\frac{1}{2}\widetilde\Eh\big(\Theta\cap[0,\pi)\times R\big)+\frac{1}{2}\widetilde\Eh\big(\big[(\Theta\cap[\pi,2\pi))-\pi\big]\times (-R)\big)
$$
where now $\Theta\in\bor{[0,2\pi)}$ and $R\in\bor\R$.

For any $T\in\trh$, let $p_{\rm ht}^T(\theta,r)$ be the density of the complex measure
$
\tr[T\Eh]
$
with respect to $(2\pi)^{-1}d\theta dr$ \cite{AlDeTo}. For example, since
$$
\<n|\Eh(\Theta\times R)|m\>=\frac{1}{2\pi}\int_\Theta e^{i(n-m)\theta}d\theta\int_R h_n(r)h_m(r)dr,
$$
it follows that 
$$
p_{\rm ht}^{|m\>\<n|}(\theta,r)=e^{i(n-m)\theta}h_n(r)h_m(r)=e^{i(n-m)\theta}p_{\rm ht}^{|m\>\<n|}(0,r).
$$
The next proposition shows that, if $T$ is regular enough, one can express $p_{\rm ht}^T$ using
the matrix elements of $T$.

\begin{proposition}\label{prop1}
For any $T\in \cal T^S(\hi)$ we have:
\begin{enumerate}
\item the double sequence $(T_{mn})_{m,n\in\N}$ is rapidly decreasing,\footnote{That is, the series
$
\sum_{m,n=0}^\infty m^{2j}n^{2k}|T_{mn}|^2
$
converge for all $j,\,k\in\N$.}
\item the integral kernel $\tilde T(x,y)=\sum_{m,n=0}^\infty T_{mn}h_m(x)h_n(y)$ where the double series converges absolutely for all $x,\,y\in\R$,\footnote{Obviously, here we have chosen a representative of the equivalence class of kernels $\tilde T$; if one chooses another representative then the results of this proposition holds only for {\it almost} all points.}
\item $p_{\rm ht}^T(0,r)=\sum_{m,n=0}^\infty T_{mn}h_n(r)h_m(r)=\tilde T(r,r)$ for (almost) all $r\in\R$,
\item the function $r\mapsto \tilde T(r,r)$ belongs to $\cal S(\R)$,
\item $e^{-i\theta N}T e^{i\theta N}\in \cal T^S(\hi)$ for all $\theta\in\R$,
\item $p_{\rm ht}^T(\theta,r)=p_{\rm ht}^{e^{-i\theta N}Te^{i\theta N}}(0,r)=\sum_{m,n=0}^\infty T_{mn}e^{i(n-m)\theta}h_n(r)h_m(r)$ for (almost) all $\theta\in[0,2\pi)$ and $r\in\R$,
\item the Wigner function $W^T\in \cal S(\R^2)$,
\item the Radon transform
$(\Ra W^T)(\theta,r)=p_{\rm ht}^T(\theta,r)$ for (almost) all $\theta\in[0,2\pi)$ and $r\in\R$.
\end{enumerate}
\end{proposition}

\begin{proof}
(1) is a standard result, see e.g.\ \cite[theorem V.13]{ReSi}.
Item (2) follows from the facts that
$$
\sup\{|h_n(x)|\ge 0\,|\,n\in\N,\,x\in\R\}<1
$$
(see, equation (22.14.17) of \cite{AbSt} in p.\ 787) and $\sum_{m,n=0}^\infty |T_{mn}|<\infty$;
moreover, for all $\psi=\sum_{n=0}^\infty c_n h_n$, $\sum_{n=0}^\infty|c_n|^2<\infty$,
$$
\int_\R \tilde T(x,y)\psi(y)dy=\sum_{m,n=0}^\infty T_{mn}h_m(x)\int_\R h_n(y)\psi(y)dy
=\sum_{m=0}^\infty h_m(x)\sum_{n=0}^\infty T_{mn}c_n=(T\psi)(x)
$$
for almost all $x\in\R$. Item (3) follows from
$$
\tr[T\Q_0(R)]=\int_R\tilde T(r,r)dr,\hspace{1cm}R\in\cal B(\R).
$$
Since, for any $f\in\cal S(\R^2)$, the mapping $r\mapsto f(r,r)$ belongs to $\cal S(\R)$, item (4) follows.
Item (5) is obvious and (6) is easy to see since $\Q_\theta(R)=e^{i\theta N}\Q_0(R)e^{-i\theta N}$ for all $R\in\bor\R$, $\theta\in\R$. Item (7) is a standard result (see, e.g.\ prop.\ 4 of \cite{AlDeTo}) and (8) is proved in \cite[prop.\ 2]{AlDeTo}.
\end{proof}


\section{Results}

In this section, we will find different connections between phase space observables $\E_K$, $K\in\sh$, and $\Eh$, the observable associated with homodyne tomography. Our results are based on the properties of the Radon transform $\Ra$ and on the Hilbert transform $\H$ which are shortly introduced in the Appendix. The main mathematical results to be used are the
{\it convolution theorem \eqref{co}, Radon inversion theorem \eqref{in}, {\em and} Plancherel formula \eqref{Pl}}.

\subsection{Convolution theorem and Radon inversion theorem}
Let $T$, $K\in\trh$.
The density of $Z\mapsto\tr[T\E_K(Z)]$ at $z_1=(q_1+ip_1)/\sqrt{2}\in\C$ is
\begin{eqnarray*}
p^T_K(z_1)&:=&
\tr{[TD(z_1)KD(z_1)^*]}=4\pi \int_\C W^T(z)W^K(z-z_1)d^2 z\\
&=&2\pi\iint_{\R^2} W^T(q,p)W^K(q-q_1,p-p_1)dqdp =2\pi\iint_{\R^2} \tilde W^K(q_1-q,p_1-p)W^T(q,p)dqdp\\
&=&2\pi(W^T*\tilde W^K)(q_1,p_1)=2\pi(\tilde W^K*W^T)(q_1,p_1)
\end{eqnarray*}
where $\tilde W^K(q,p):=W^K(-q,-p)$. If $W^T$ and $W^K$ are integrable then it follows from proposition 2 of \cite{AlDeTo} that
$$
(\Ra W^T)(\theta,r)=p_{\rm ht}^T(\theta,r)
$$
for $d\theta dr$-almost all $(\theta,r)$, and similarly for $W^K$. Moreover,
 one gets from the convolution theorem \eqref{co} that
\begin{equation}\label{tulos}
({\Ra}p^T_K)(\theta,r)=2\pi\int_\R p^T_{\rm ht}(\theta,s)p^K_{\rm ht}(\theta,s-r)ds
\end{equation}
for $d\theta dr$-almost all $(\theta,r)$.
Note that, {\it formally}, replacing $p_K^T$ with $W^T=\frac{1}{2\pi}p^T_{2\Pi}$ (which is only a quasiprobability distribution for non-Gaussian states) in the above equation, we get
$$
({\Ra}W^T)(\theta,r)=p_{\rm ht}^T(\theta,r)=\int_\R p^T_{\rm ht}(\theta,s)\delta(s-r)ds
$$
and $p^{2\Pi}_{\rm ht}(\theta,r)=\delta(r)$
where $\delta(s-r)ds$ is the Dirac measure concentrated on $r$, so that, the phase space $p^T_K$ distribution is a kind of a 'smoothed version' of quadature data (associated with $W^T$).

Keeping equation \eqref{tulos} in mind, one gets from Radon inversion theorem \eqref{in} that
\begin{equation}\label{tulos2}
p_K^T=\frac{1}{2\pi}\Ra^*\Lambda\underbrace{(\Ra p^T_K)}_{\text{$p^T_{\rm ht}$, $p^K_{\rm ht}$}}
\end{equation}
if $p_K^T$ is a Scwartz function.\footnote{This holds, e.g., when $T,\,K\in\cal T^S(\hi)$ since then $p^T_K$ is a convolution of Schwartz functions $W^T$ and $\tilde W^K$ and thus a Schwartz function (see, proposition \ref{prop1}).}
Thus, the probability distribution $p_K^T$ where $T,\,K\in\cal T^S(\hi)\cap \sh$ can be constructed by using homodyne tomography data for both $T$ and $K$.
Physically, this means that to {\it reconstruct} the measurement outcome statistics $p^T_K$ of an 8-port homodyne detector, one needs to measure the statistics $p^T_{\rm ht}$ and $p^K_{\rm ht}$ separately by using balanced homodyne detection for quandratures, and then use formulas \eqref{tulos} and \eqref{tulos2}.
In some cases, this could be easier in the sense that, if one needs to prepare or measure the kernel state $K$, one can use quadrature measurements (tomography) to get $p^K_{\rm ht}$ (and find the state $K$ if it is unknown a priori).
Then $p^T_K$ can be obtained directly either by using 8-port detection (since $K$ is known) or by measuring $p^T_{\rm ht}$ via quantum tomography.
Since that $p^T_ K$ is (almost) symmetric in the interchange of $K$ and $T$, that is, $p^T_K(z_1)=p^K_T(-z_1),$
{\it mathematically,} the signal state $T$ and the kernel state $K$ are in the same role in 8-port homodyne detection.

\subsection{Plancherel formula and Markov kernels}
Let $K\in\trh$ be such that $W^K\in L^1(\R^2)$ and
$(q,p)\in\R^2$, $(q_1,p_1)\in\R^2$, and $\theta\in[0,2\pi)$.
If one denotes $W_{q_1,p_1}^K(q,p):=W^K(q-q_1,p-p_1)$ 
and defines parameters $a:=q_1\cos\theta+p_1\sin\theta$, $b:=-q_1\sin\theta+p_1\cos\theta$,
one gets
\begin{eqnarray*}
(\Ra W_{q_1,p_1}^K)(\theta,r)
&=&
\int_\R W_{q_1,p_1}^K(r\cos\theta-t\sin\theta,\,r\sin\theta+t\cos\theta)dt \\
&=&
\int_\R W^K(r\cos\theta-t\sin\theta-q_1,\,r\sin\theta+t\cos\theta-q_2)dt \\
&=&
\int_\R W^K\big((r-a)\cos\theta-(t-b)\sin\theta,\,(r-a)\sin\theta+(t-b)\cos\theta\big)d(t-b) \\
&=&
(\Ra W^K)(\theta,r-a)=(\Ra W^K)(\theta,r-q_1\cos\theta-p_1\sin\theta) \\
&=&p^K_{\rm ht}(\theta,r-q_1\cos\theta-p_1\sin\theta).
\end{eqnarray*}
Assume then that $K,\,T\in\cal T^S(\hi)$ and define
$M^K_{q_1,p_1}:=\Lambda (\Ra W_{q_1,p_1}^K)$ (see, equation \eqref{lambda} in the Appendix).
Since $$p^T_K(q_1,p_1)=2\pi \iint_{\R^2} W^T(q,p)W^K(q-q_1,p-p_1)dqdp$$ we get
from Plancherel formula \eqref{Pl} that
$$
p_K^T(q_1,p_1)=\frac{1}{2\pi}
\int_0^{2\pi}\int_\R
p^T_{\rm ht}(\theta,r)M^K_{q_1,p_1}(\theta,r) d\theta dr.
$$
Hence, for any $Z\in\bor{\R^2}$,
$$
\tr[T\E_K(Z)]=\frac{1}{2\pi}\int_Z p_K^T(q_1,p_1)dq_1dp_1=\frac{1}{2\pi}
\int_Z\int_0^{2\pi}\int_\R M^K_{q_1,p_1}(\theta,r)d\tr[T\Eh(\theta,r)]dq_1dp_1
$$
If $Z$ is compact, it follows from Fubini's theorem, that one can change the order of the above integrations and get for operator measures $\E_K$ and $\Eh$ (by the density of $\cal T^S(\hi)\subset\trh$) that
\begin{eqnarray*}
\boxed{
\E_K(Z)=
\int_0^{2\pi}\int_\R\underbrace{\left[
\frac{1}{2\pi}\int _Z M^K_{q_1,p_1}(\theta,r)dq_1dp_1
\right]}_{=:\,{\cal M_K(Z;\theta,r)}
}
d{\E_{\rm ht}}(\theta,r)
=\int_0^{2\pi}\int_\R
{\cal M_K(Z;\theta,r)}
d{\E_{\rm ht}}(\theta,r)
}
\end{eqnarray*}
(weakly) for all compact $Z\subset\R^2$,
where $\cal M_K$ is a {\it generalized Markov kernel}.\footnote{Note that the effects $\E_K(Z)$ of compact sets $Z$ fully determine the POM $\E_K$, $K\in \cal T^S(\hi)\cap\sh$.}
Since $\cal M_K$ is not necessarily a nonnegative function (see examples 1 and 2),
 $\cal M_K(Z;\theta,r)$ is not true conditional or transition probability.
However, one could say that the measurement $\E_K$ is subordinate to the measurement $\Eh$ (see, \cite{Ho})
or $\E_K$ is a postprocessing of $\Eh$.

Since $p_K^T(q_1,p_1)=\sum_{m,n=0}^\infty K_{mn} p_{\kb m n}^T(q_1,p_1)$ for all $T\in\trh$ (where the double series converges absolutely) it follows that (weakly)
\begin{eqnarray*}
\int_0^{2\pi}\int_\R
{\cal M_K(Z;\theta,r)}
d{\E_{\rm ht}}(\theta,r) &=& \E_K(Z)
=\sum_{m,n=0}^\infty K_{mn}\E_{\kb m n}(Z)\\
&=&\sum_{m,n=0}^\infty K_{mn}
\int_0^{2\pi}\int_\R
{\cal M_{\kb m n}(Z;\theta,r)}
d{\E_{\rm ht}}(\theta,r)
\end{eqnarray*}
for all compact $Z\subset\R^2$. Thus, one needs to find the density $M^K_{q_1,p_1}(\theta,r)$ of the Markov kernel only in the case of $K=|m\>\<n|$ where $m,\,n\in\N$.
Before doing that we study some properties of the densities which are immediate from the definitions.

We notice that, when $q_1=0=p_1$, one gets $p_K^T(0,0)=\tr{[TK]}$ where $T,\,K\in\trh$.
Thus, if $T,\,K\in\cal T^S(\hi)$,
$$
\frac{1}{2\pi}
\int_0^{2\pi}\int_\R
M^K_{0,0}(\theta,r) p^T_{\rm ht}(\theta,r)d\theta dr=\tr{[TK]}.
$$
Especially, by choosing $K=|m\>\<n|$, $m,\,n\in\N$, one recovers the matrix element $T_{nm}$ from the tomographic data
$p^T_{\rm ht}$ as follows:
$$
T_{nm}=\frac{1}{2\pi}
\int_0^{2\pi}\int_\R
M^{\kb m n}_{0,0}(\theta,r) p^T_{\rm ht}(\theta,r)d\theta dr
$$
(see, section 5 of \cite{Le} and references therein). 
If $T=\kb k l$, $k,\,l\in\N$, then
$$
K_{lk}=\frac{1}{2\pi}
\int_0^{2\pi}\int_\R
M^K_{0,0}(\theta,r) p^{\kb kl}_{\rm ht}(\theta,r)d\theta dr
=\frac{1}{2\pi}
\int_0^{2\pi}\int_\R
M^K_{0,0}(\theta,r) e^{i(l-k)\theta}h_l(r)h_k(r)d\theta dr
$$
since $p_{\rm ht}^{|k\>\<l|}(\theta,r)=e^{i(l-k)\theta}h_l(r)h_k(r)$.
Combining the above two results, we get the {\it orthogonality condition:}
\begin{equation}\label{orto}
\frac{1}{2\pi}
\int_0^{2\pi}\int_\R
M^{\kb m n}_{0,0}(\theta,r) e^{i(l-k)\theta}h_l(r)h_k(r)d\theta dr=\delta_{m,l}\delta_{n,k}
\end{equation}
(where $\delta_{m,l}$ is the Kronecker delta). 
Next we calculate the structure of $M^K_{q_1,p_1}(\theta,r)$ in the case of $K=|m\>\<n|$.

Recall that $M^K_{q_1,p_1}=\Lambda (\Ra W_{q_1,p_1}^K)$ where 
$(\Ra W_{q_1,p_1}^K)(\theta,r)=p^K_{\rm ht}(\theta,r-a)$ and 
$
a=q_1\cos\theta+p_1\sin\theta. 
$
Since $p_{\rm ht}^{|m\>\<n|}(\theta,r)=e^{i(n-m)\theta}h_n(r)h_m(r)$
one gets 
$$
(\Ra W^{|m\>\<n|}_{q_1,p_1})(\theta,r)=e^{i(n-m)\theta}h_n(r-a)h_m(r-a).
$$
Since the Hilbert transform $\H$ is shift invariant and commutes with derivation (see Appendix)
one needs only to find $\H(h_nh_m)$.

Consider first the case $n=m=0$ (i.e.\ the Husimi $Q$-function case where $K=|0\>\<0|$). Then
$$
h_0(x) = \frac{1}{\sqrt[4]{\pi}}e^{-\frac 12 x^2} \; \text{ and } \;
h_0(x)^2 = \frac{1}{\sqrt{\pi}}e^{-x^2}.
$$
Let ${\rm daw}(r):=e^{-r^2}\int_0^r e^{x^2}dx$ be the {\it Dawson's integral}
which is related to the imaginary error function erfi in the obvious way (see, e.g.\ \cite{AbSt, SpOl}).
The result of the next lemma is well-known but, for completeness, we give an alternative proof for it.

\begin{lemma} \label{lemma1}
For all $r\in\R$,
$$
[\H(h_0^2)](r)=\frac{2}{\pi}e^{-r^2}\int_0^r e^{x^2}dx
=\frac{2}{\pi}{\rm daw}(r)=\frac{1}{\sqrt{\pi}}e^{-r^2}{\rm erfi}(r).
$$
\end{lemma}

\begin{proof}
Let
$$
f(r):=[\H(h_0^2)](r)=\frac{1}{\pi^{3/2}}\lim_{\epsilon\to0+}\int_{x\in\R\atop|r-x|>\epsilon}\frac{e^{-x^2}}{r-x}dx.
$$
Since 
\begin{eqnarray*}
\frac{d f(r)}{d r}
&=&
\left[\H\left(\frac{d h_0^2}{d x}\right)\right](r)=
\frac{1}{\pi^{3/2}}\lim_{\epsilon\to0+}\int_{x\in\R\atop|r-x|>\epsilon}\frac{-2x e^{-x^2}}{r-x}dx \\
&=&
\frac{2}{\pi^{3/2}}\left[\lim_{\epsilon\to0+}\int_{x\in\R\atop|r-x|>\epsilon}\frac{(r-x) e^{-x^2}}{r-x}dx
-\lim_{\epsilon\to0+}\int_{x\in\R\atop|r-x|>\epsilon}\frac{r e^{-x^2}}{r-x}dx\right]\\
&=&
\frac{2}{\pi^{3/2}}\left[\int_\R e^{-x^2}dx
-r\,\lim_{\epsilon\to0+}\int_{x\in\R\atop|r-x|>\epsilon}\frac{e^{-x^2}}{r-x}dx\right]=\frac{2}{\pi}-2 r f(r)
\end{eqnarray*}
and 
$$
\frac{d}{dr}\frac{2}{\pi}{\rm daw}(r)=\frac{2}{\pi}-2 r \frac{2}{\pi}{\rm daw}(r)
$$
it follows that
$$
f(r)=\frac{2}{\pi}{\rm daw}(r)+ce^{-r^2}
$$
where $c\in\R$.
But 
$$
f(0)=\frac{1}{\pi^{3/2}}\lim_{\epsilon\to0+}\int_{x\in\R\atop|x|>\epsilon}\frac{e^{-x^2}}{-x}dx
=\frac{1}{\pi^{3/2}}\lim_{\epsilon\to0+}\int_\epsilon^\infty\frac{e^{-(- y)^2}-e^{- y^2}}{ y}d y=0
=0+c
$$
so that $c=0$ and the lemma follows.
\end{proof}

Since
$
(\Ra W^{|0\>\<0|}_{q_1,p_1})(\theta,r)=[h_0(r-a)]^2,
$
$
(\Lambda\fii)(\theta,r)=\pi \frac{d}{d r}(\H\fii_\theta)(r),
$
and
$
M^{\kb00}_{q_1,p_1}=\Lambda (\Ra W^{\kb00}_{q_1,p_1}),
$
the above lemma implies that
$$
M^{|0\>\<0|}_{q_1,p_1}(\theta,r)
=2\frac{\partial}{\partial r}{\rm daw}(r-q_1\cos\theta-p_1\sin\theta)
=\sum_{k=0}^\infty \frac{(-1)^k k!}{2^k(2k)!}H_{2k}(r-q_1\cos\theta-p_1\sin\theta)
$$
is an analytic function vanishing at infinity $r\to\pm\infty$ (see, lemma 1 of \cite{KiLaPe}).
Moreover,
$$
\E_{|0\>\<0|}(Z)=
\int_0^{2\pi}\int_\R
\left[\frac{1}{2\pi}
\int _Z M^{|0\>\<0|}_{q_1,p_1}(\theta,r)dq_1dp_1
\right]
d{\E_{\rm ht}}(\theta,r)
$$
(see a direct verification from \cite{LaPe}).
Next we consider the general case when $K=\kb m n$ where $m,\,n\in\N$.

\begin{proposition}For all $r\in\R$,
$$
[\H(h_n h_m)](r)=\frac{(-1)^{n+m}}\pi \sqrt{\frac{n!m!}{2^{n+m}}}
\sum_{v=0}^{\min\{n,m\}}
\frac{2^v}{v!(n-v)!(m-v)!}\frac{d^{n+m-2v}(2{\rm daw}(r))}{d r^{n+m-2v}}
$$
\end{proposition}

\begin{proof}
By the {\it Feldheim's identity} \cite[eq.\ (1.4)]{Fe}
$$
H_n(r)H_m(r)=n! m! \sum_{v=0}^{\min\{n,m\}}\frac{2^v}{v!(n-v)!(m-v)!}H_{n+m-2 v}(r)
$$
and 
{\it Rodrigue's formula} 
$$
H_k(r)=(-1)^k e^{r^2}{d^k e^{-r^2}\over dr^k}
$$
one can write
\begin{eqnarray*}
h_n(r)h_m(r) &=& \frac{1}{\sqrt{2^{n+m}n! m! \pi}}H_n(r)H_m(r)e^{-r^2}  \\
&=& \frac{n! m!}{\sqrt{2^{n+m}n! m! \pi}} \sum_{v=0}^{\min\{n,m\}}\frac{2^v}{v!(n-v)!(m-v)!}H_{n+m-2 v}(r)e^{-r^2} \\
&=& \frac{1}{\sqrt\pi}\sqrt{\frac{n! m!}{2^{n+m}}} \sum_{v=0}^{\min\{n,m\}}\frac{2^v}{v!(n-v)!(m-v)!}(-1)^{n+m-2 v} {d^{n+m-2 v} e^{-r^2}\over dr^{n+m-2 v}}.
\end{eqnarray*}
Since $e^{-r^2}=\sqrt\pi \, h_0(r)^2$ and $\H$ commutes with derivation, one gets
$$
\H(h_n h_m)=(-1)^{n+m}
\sqrt{\frac{n! m!}{2^{n+m}}} \sum_{v=0}^{\min\{n,m\}}\frac{2^v}{v!(n-v)!(m-v)!} {d^{n+m-2 v} \H(h_0^2)
\over dr^{n+m-2 v}}
$$
and the proposition follows from lemma \ref{lemma1}.
\end{proof}

For any smooth function $f:\,\R\to\R$ we let $f^{(s)}$ denote the $s$th derivative function of $f$ (and $f^{(0)}:=f$).
Let 
$
Y:=2\,{\rm daw}^{(1)}.
$
As shown in the Appendix of \cite{KiLaPe}, for all $p\in\N$,
\begin{eqnarray} \label{even}
Y^{(2p)}(r) &=& (-1)^p 2^p \sum_{k=0}^\infty \frac{(-1)^k (k+p)!}{2^k(2k)!}H_{2k}(r), \\
Y^{(2p+1)}(r) &=& (-1)^{p+1} 2^p \sum_{k=0}^\infty \frac{(-1)^k (k+p+1)!}{2^k(2k+1)!}H_{2k+1}(r) \label{odd}
\end{eqnarray}
for all $r\in\R$. Finally, we are ready to calculate the density $M^{\kb m n}_{q_1,p_1}$ of the kernel associated with $\kb m n$: from the previous proposition one gets
\begin{eqnarray}\nonumber
M^{\kb m n}_{q_1,p_1}(\theta,r)&=&\pi e^{i(n-m)\theta}\frac{\partial [\H(h_n h_m)](r-a)}{\partial r}\\
&=&e^{i(n-m)\theta}(-1)^{n+m} \sqrt{\frac{n!m!}{2^{n+m}}}
\sum_{v=0}^{\min\{n,m\}}
\frac{2^v}{v!(n-v)!(m-v)!}Y^{(n+m-2v)}(r-a)  \label{kernel}
\end{eqnarray}
where $a=q_1\cos\theta+p_1\sin\theta$.
Immediately one sees that $\R\ni r\mapsto M^{\kb m n}_{0,0}(0,r)\in\R$ is analytic and vanishes at infinity $r\to\pm\infty$.
Moreover,
\begin{eqnarray*}
M^{\kb m n}_{q_1,p_1}(\theta,r)&=&e^{i(n-m)\theta}M^{\kb m n}_{0,0}(0,r-q_1\cos\theta-p_1\sin\theta)=\overline{M^{\kb n m}_{q_1,p_1}(\theta,r)}
\end{eqnarray*}
and the orthogonality relation \eqref{orto} reduces to
$$
\int_\R
M^{\kb m n}_{0,0}(0,r) h_l(r)h_k(r)dr 
\underbrace{\frac{1}{2\pi}
\int_0^{2\pi} e^{i(l-m-k+n)\theta}d\theta}_{=\,\delta_{l-m-k+n,0}}=\delta_{m,l}\delta_{n,k}.
$$
Obviously,
without restricting generality, we may assume that $\theta=0$, $q_1=0$, and $p_1=0$  in the next theorem:

\begin{theorem}\label{theorem1}
If $n+m$ is even, then
$$
M^{\kb m n}_{0,0}(0,r)=\frac{(-1)^{(n+m)/2}}{\sqrt{n!\,m!}}\sum_{k=(n+m)/2}^\infty
\frac{(-1)^k}{2^k(2k)!}
\frac{\big(k+\frac{1}{2}(n-m)\big)!\,\big(k+\frac{1}{2}(m-n)\big)!}{\big(k-\frac{1}{2}(n+m)\big)!}H_{2k}(r),
$$
and if $n+m$ is odd, then
$$
M^{\kb m n}_{0,0}(0,r)
=\frac{(-1)^{(n+m-1)/2}}
{\sqrt{2 \, n!m!}} \sum_{k={(n+m-1)/2}}^\infty \frac{(-1)^k }{2^k(2k+1)!}
\frac{\big(k+\frac12{(n-m+1)}\big)!\,\big(k+\frac12{(m-n+1)}\big)!}{\big(k-\frac12{(n+m-1)}\big)!}H_{2k+1}(r)
$$
for all $r\in\R$.
\end{theorem}

\begin{proof}
Suppose that $n+m$ is even so that one can write $n+m=2l$ where $l\in\N$. Then from equations \eqref{kernel} and \eqref{even} one gets
\begin{eqnarray*}
&& M^{\kb m n}_{0,0}(0,r)
=(-1)^{n+m} \sqrt{\frac{n!m!}{2^{n+m}}}
\sum_{v=0}^{\min\{n,m\}}
\frac{2^v}{v!(n-v)!(m-v)!}Y^{(n+m-2v)}(r) \\
&&
=\frac{\sqrt{n!m!}}{2^l}
\sum_{v=0}^{\min\{n,m\}}
\frac{2^v}{v!(n-v)!(m-v)!}
(-1)^{l-v} 2^{l-v} \sum_{k=0}^\infty \frac{(-1)^k (k+l-v)!}{2^k(2k)!}H_{2k}(r) \\
&&
=\sqrt{n!m!}\sum_{k=0}^\infty \frac{(-1)^k}{2^k(2k)!}H_{2k}(r)
\sum_{v=0}^{\min\{n,m\}}
\frac{ (-1)^{l-v}  (k+l-v)!}{v!(n-v)!(m-v)!} \\
&&
=\sqrt{n!m!}(-1)^l\sum_{k=0}^\infty \frac{(-1)^k}{2^k(2k)!}H_{2k}(r)
\frac{(k+l-m)!}{n!}
\underbrace{\sum_{v=0}^{\min\{n,m\}} (-1)^{v}
{n\choose v}
{k+l-v\choose m-v}}_{=\, {k+l-n \choose m}\;\text{ (see eq.\ (5) of \cite{Riordan} in p.\ 8)}} \\
&&
=\sqrt{n!m!}(-1)^l\sum_{k=l}^\infty \frac{(-1)^k}{2^k(2k)!}H_{2k}(r)
\frac{(k+l-m)!(k+l-n)!}{n!m!(k+l-n-m)!}\\
&&
=\frac{(-1)^l}{\sqrt{n!m!}}\sum_{k=l}^\infty \frac{(-1)^k}{2^k(2k)!}\frac{(k+l-m)!(k+l-n)!}{(k+l-n-m)!}H_{2k}(r)
\end{eqnarray*}
and the first equation follows by substituting $l=(n+m)/2$ into the above equation.

Similarly, if $n+m=2 l + 1$, $l\in\N$, then from \eqref{kernel} and \eqref{odd} it follows that
\begin{eqnarray*}
&& M^{\kb m n}_{0,0}(0,r)
=(-1)^{n+m} \sqrt{\frac{n!m!}{2^{n+m}}}
\sum_{v=0}^{\min\{n,m\}}
\frac{2^v}{v!(n-v)!(m-v)!}Y^{(n+m-2v)}(r) \\
&&
=\sqrt{\frac{n!m!}{2}} \sum_{k=0}^\infty \frac{(-1)^k }{2^k(2k+1)!}H_{2k+1}(r)
\sum_{v=0}^{\min\{n,m\}}
\frac{(-1)^{l-v}(k+l-v+1)!}{v!(n-v)!(m-v)!} \\
&&
=\sqrt{\frac{n!m!}{2}} (-1)^l\sum_{k=0}^\infty \frac{(-1)^k }{2^k(2k+1)!}H_{2k+1}(r)
\frac{(k+1+l-m)!}{n!}{k+1+l-n \choose m} \\
&&
=\frac{(-1)^l}{\sqrt{2 \, n!m!}} \sum_{k=l}^\infty \frac{(-1)^k }{2^k(2k+1)!}
\frac{(k+1+l-m)!(k+1+l-n)!}{(k+1+l-n-m)!}H_{2k+1}(r)
\end{eqnarray*}
and the second equation has been proved.
\end{proof}

\begin{remark}
Since $Y$ is an analytic function (see, eq.\ (A.6) of \cite{KiLaPe}) one sees from equation \eqref{kernel} that
$
r\mapsto M^{\kb m n}_{0,0}(0,r)
$
is an analytic function. Indeed, its Maclaurin series is easy to compute by using the series
\begin{equation}\label{Y}
Y(r)=2\sum_{u=0}^\infty\frac{(-1)^u u!}{(2u)!}(2r)^{2u}
\end{equation}
which implies that, for all $p\in\N$,
$$
Y^{(p)}(r)=2\sum_{u\in\N\atop u\ge p/2}\frac{(-1)^u u!}{(2u-p)!}2^{2u} r^{2u-p}.
$$
For example, if $n+m=2l$, $l\in\N$, one gets
\begin{eqnarray*}
M^{\kb m n}_{0,0}(0,r)
&=&(-1)^{n+m} \sqrt{\frac{n!m!}{2^{n+m}}}
\sum_{v=0}^{\min\{n,m\}}
\frac{2^v}{v!(n-v)!(m-v)!}Y^{(n+m-2v)}(r)  \\
&=& \sqrt{\frac{n!m!}{2^{2l}}}
\sum_{v=0}^{\min\{n,m\}}
\frac{2^v}{v!(n-v)!(m-v)!}
2\sum_{u=l-v}^\infty\frac{(-1)^u u!}{(2u-2l+2v)!}2^{2u} r^{2(u-l+v)} \\
&=\atop{t=u-l+v}&2 \sqrt{n!m!}\sum_{t=0}^\infty\frac{(-1)^{t}(2r)^{2t}}{(2t)!}
\sum_{v=0}^{\min\{n,m\}}
\frac{(-2)^{l-v}({l-v+t})!}{v!(n-v)!(m-v)!}.
\end{eqnarray*}
\end{remark}

\subsection{Examples}

\begin{example}
Let us consider the case when $K=\kb\psi\psi$ where $\psi$ is a Schrödinger cat state:
$$
\psi=\frac{1}{\sqrt2}\big(\ket 0+\ket 1\big).
$$
From \eqref{kernel} we get $M^{\kb00}_{0,0}(\theta,r)=Y(r)$,
$M^{\kb01}_{0,0}(\theta,r)=-e^{i\theta}Y^{(1)}(r)/\sqrt{2}$,
and $M^{\kb11}_{0,0}(\theta,r)=Y(r)+Y^{(2)}(r)/2$, so that
\begin{eqnarray*}
M^{\kb\psi\psi}_{0,0}(\theta,r)&=&Y(r)-\frac1{\sqrt 2}Y^{(1)}(r)\cos\theta+\frac{1}{4}Y^{(2)}(r) \\
&=&2(r+\sqrt 2\cos\theta)\big[r+(1-2r^2){\rm daw}(r)\big].
\end{eqnarray*}
The function $[0,\pi)\times\R\ni (\theta,r)\mapsto M^{\kb\psi\psi}_{0,0}(\theta,r)\in\R$ is plotted in the next picture by interpreting 
$(\theta,r)$ as the polar coordinates of $\R^2$.

\begin{figure}[tbh]
\subfigure[]{\label{fig1a}\includegraphics[width=0.45\columnwidth]{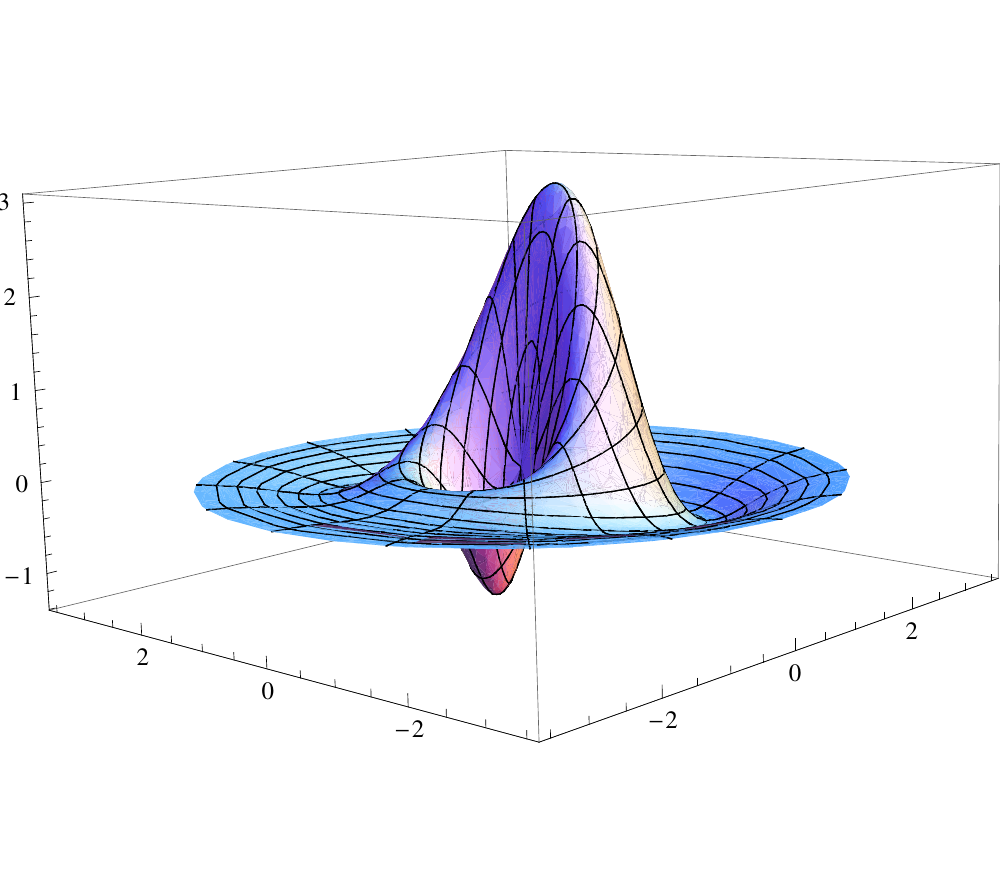}}
\quad
\subfigure[]{\label{fig1b}\includegraphics[width=0.45\columnwidth]{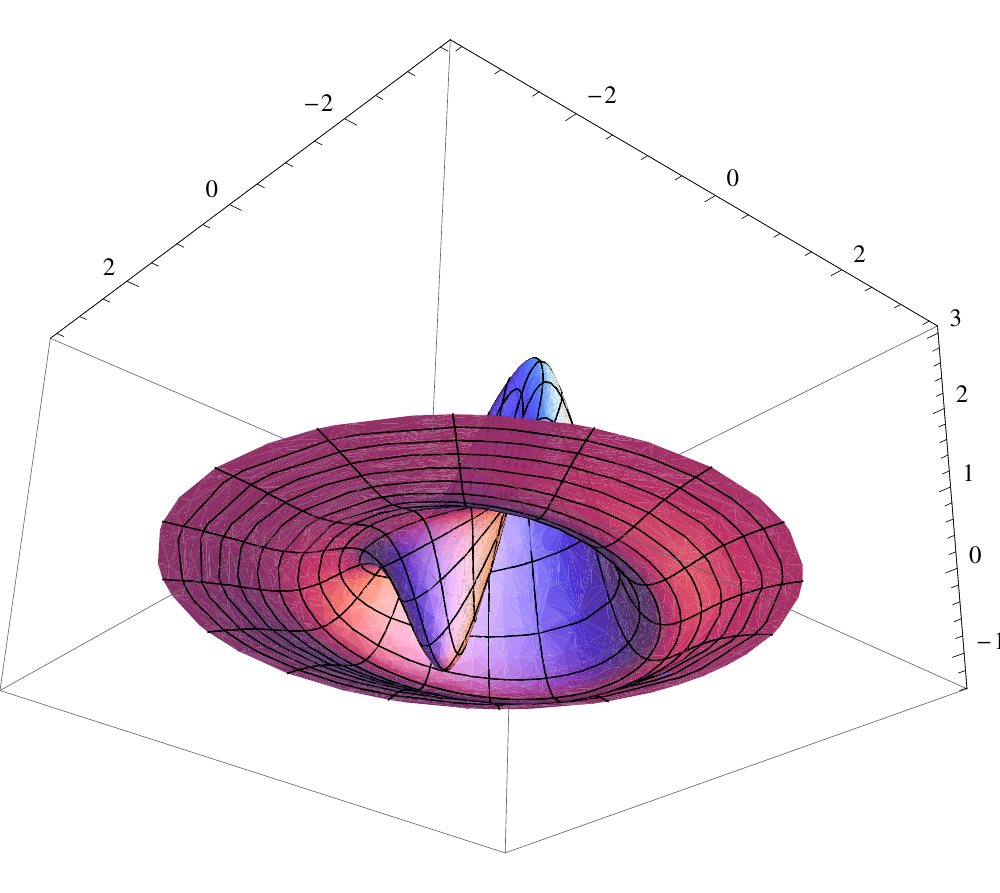}}
\caption{The density $M^{\kb\psi\psi}_{0,0}(\theta,r)$ of the generalized Markov kernel in the case of a Schrödinger cat state $\psi=\frac{1}{\sqrt2}\big(\ket 0+\ket 1\big).$}
\end{figure}

\end{example}

\begin{example}In this example, we study the Cahill-Glauber $s$-parametrized quasiprobability distributions \cite{CaGl}.
Instead of using $s$ as a parameter, we define $\lambda:=(s+1)/(s-1)$ and assume that $\lambda\in[-1,1)$ (that is,
$s=(\lambda+1)/(\lambda-1)\in(-\infty,0]$).
Let $T\in\trh$.
Now the $s$-distribution is $p_{K_\lambda}^T$ where the kernel
$K_\lambda:=(1-\lambda)\sum_{n=0}^\infty\lambda^n\kb n n$ \cite[eq.\ (6.23)]{CaGl}.
Note that $K_\lambda\in\cal T^S(\hi)$ only when $\lambda\in(-1,1)$, and then $\tr{K_\lambda}=1$.
For $\lambda=0$ (and $s=-1$), $K_0=\kb 00$, and $p^T_{K_0}$ is the Husimi $Q$-function.
When $\lambda=-1$ (and $s=0$) we have $K_{-1}=2\Pi$ and
$
p^T_{K_{-1}}=p^T_{2\Pi}=2\pi W^T.
$
Note that, for $\lambda\in(-1,1)$, one can define an operator measure $\E_{K_\lambda}$
and calculate $M^{K_\lambda}_{0,0}(\theta,r)$ as before, but when 
$\lambda=-1$ the corresponding $\E_{K_{-1}}$ is only a generalized operator measure, that is, a sesquilinear form valued measure \cite{Pe}. Hence, we assume that $\lambda\in(-1,1)$ and finally take the limit $\lambda\to-1$.

By theorem \ref{theorem1}, one sees that
\begin{eqnarray*}
M^{K_\lambda}_{0,0}(\theta,r)&=&(1-\lambda)\sum_{n=0}^\infty \lambda^n
M^{\kb n n}_{0,0}(\theta,r)=
(1-\lambda)\sum_{n=0}^\infty 
\frac{(-\lambda)^{n}}{n!}\sum_{k=n}^\infty
\frac{(-1)^k}{2^k(2k)!}
\frac{(k!)^2}{(k-n)!}H_{2k}(r) \\
&=&(1-\lambda)
\sum_{k=0}^\infty
\frac{(-1)^k k!}{2^k(2k)!}
H_{2k}(r) \underbrace{\sum_{n=0}^k
\frac{(-\lambda)^{n}k!}{n!(k-n)!}}_{=\,(1-\lambda)^k}=(1-\lambda)
\sum_{k=0}^\infty
\frac{(\lambda-1)^k k!}{2^k(2k)!}
H_{2k}(r).
\end{eqnarray*}
It is easy to calculate the Maclaurin series of $r\mapsto M^{K_\lambda}_{0,0}(\theta,r)$ (see eq.\ (A.1) and the text above eq.\ (A.6) of \cite{KiLaPe}): we get
$$
M^{K_\lambda}_{0,0}(\theta,r)=2\sum_{u=0}^\infty \frac{(-1)^u u! (2 r)^{2u}}{(2 u)!}\bigg(\underbrace{\frac{1-\lambda}{1+\lambda}}_{=\,-1/s}\bigg)^{u+1}.
$$
From equation \eqref{Y} one sees that
$$
M^{K_\lambda}_{0,0}(\theta,r)=-\frac1s Y(r/\sqrt{-s})=-\frac1s M^{\kb00}_{0,0}(\theta,r/\sqrt{-s}).
$$
When $\lambda\to-1$, $s\to 0$,
$$
M^{K_\lambda}_{0,0}(\theta,0)=-\frac{1}{s}Y(0)=\frac{2}{-s}\to\infty
$$
so that the density of the Markov kernel is not defined in the Wigner function limit, as expected.
Note that $M^{K_\lambda}_{0,0}$ does not define the Dirac $\delta$-distribution in the limit (even though it is an increasing peak, see the next picture), since $\int_\R Y(r)dr=0$ implies that
$$
\int_\R M^{K_\lambda}_{0,0}(\theta,r)dr=0
$$
for all $\lambda\in(-1,1)$. However, for any $(q_1,p_1)\in\R^2$ one gets
$$
2\pi W^T(q_1,p_1)=p^T_{K_{-1}}(q_1,p_1)=\lim_{\lambda\to-1+}
\frac{1}{2\pi}
\int_0^{2\pi}\int_\R
M^{K_\lambda}_{q_1,p_1}(\theta,r) p^T_{\rm ht}(\theta,r) d\theta dr
$$
but cannot change the order of taking the limit and integration.

\begin{figure}[tbh]
\includegraphics{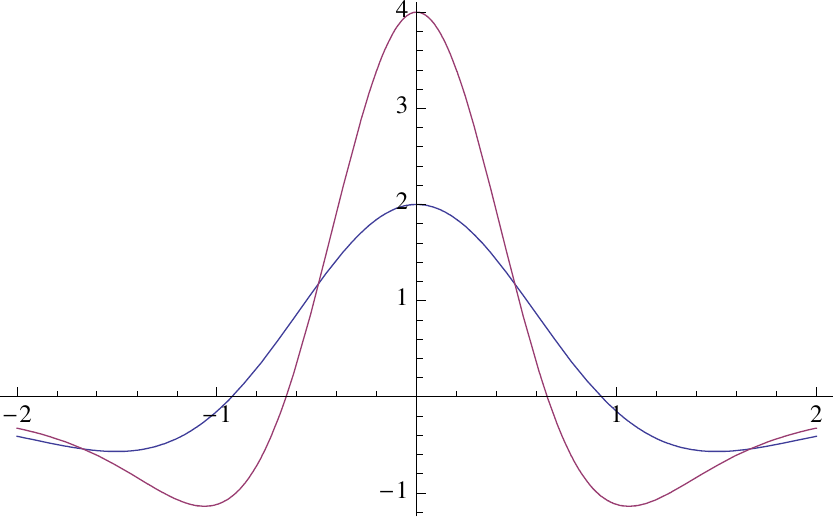} 
\caption{The density $r\mapsto M^{K_{\lambda}}_{0,0}(\theta,r)$ of the generalized Markov kernel (for all $\theta\in[0,2\pi)$). The higher curve corresponds to the case $s=-\frac12$ and for the lower curve $s=-1$.}
\end{figure}

\end{example}

\begin{example}In this example, we show that the generalized Markov kernel does not exist for all states.
Choose ${K_\eta}=|\eta\>\<\eta|$ where $\eta=\chi_{[0,1]}\in L^2(\R)$ (the characteristic function of $[0,1]$) is a unit vector.
The first observations are that ${K_\eta}\notin\cal T^S(\hi)$ and the Wigner function $W^{K_\eta}$ of ${K_\eta}$ is not integrable. Thus,
the equation $(\Ra W^{K_\eta})(\theta,r)=p_{\rm ht}^{K_\eta}(\theta,r)$ is not necessarily valid.
However, we can calculate $p_{\rm ht}^{K_\eta}(\theta,r)$. For example,
$p_{\rm ht}^{K_\eta}(0,r)=[\chi_{[0,1]}(r)]^2=\chi_{[0,1]}(r)$.
Although the Radon inversion theorem and Plancherel formula obviously fail, we can try to derive $M^{K_\eta}_{q_1,p_1}(\theta,r)$ as before we did for $\cal T^S(\hi)$-kernels. But now $d\chi_{[0,1]}(x)/dx=\delta(x)-\delta(x-1)$ must be understood as a (tempered) distribution and a formal calculation shows that
$$
M^{K_\eta}_{0,0}(0,r)=\pi\left(\H{d\chi_{[0,1]}\over dx}\right)(r)=\frac1r-\frac1{r-1}
$$
which is singular at $r=0$ and $r=1$.
On the other hand,
$$
\pi(\H\chi_{[0,1]})(r)=\ln\left|\frac{r}{r-1}\right|
$$
and a we get the same result:
$$
M^{K_\eta}_{0,0}(0,r)=\frac{d}{dr}\ln\left|\frac{r}{r-1}\right|=\frac1r-\frac1{r-1}.
$$

It is interesting to note that $\E_{K_\eta}$ is not informationally complete, since the support of
$$
(q,p)\mapsto\tr[{K_\eta}D(q,p)]=\<\eta|D(q,p)\eta\>=e^{-iqp/2}\int_0^1e^{ipx}\chi_{[0,1]}(x-q)dx
$$ 
is not the whole $\R^2$ \cite{Ali, KiWe}.
It is an open question how the informationally completeness of a phase space observable $\E_K$ 
is connected with the existence of the generalized Markov kernel $\cal M_K$.
\end{example}

\section*{Appendix: Radon and Hilbert transforms}
The basic properties of the Radon and Hilbert transforms, $\Ra$ and $\H$, respectively, are collected in this Appendix.
There is plenty of literature on these transforms. For example, \cite{He} and \cite{Ep} essentially contain the results for $\Ra$ whereas \cite{Bu} and \cite{Pa} are good references for $\H$. The elementary results of this Appendix which are easy to prove (e.g., by changing integration variables and by using Fubini's theorem), are denoted by $(\star)$.

Let $\P$ be the set of straight lines in $\R^2$. Any line $\xi(\theta,r):=\{(x,y)\in\R^2\,|\,x\cos\theta+y\sin\theta=r\}$
can be parametrized with two real numbers $\theta\in[0,2\pi)$ and $r\in\R$, and $\P$ becomes a smooth two-dimensional manifold. Since $\xi(-\theta,-r)=\xi(\theta,r)$, $\P$ has a double covering $\S\times\R\to\P$ where
$\S$ is the unit circle $\{(\cos\theta,\sin\theta)\in\R^2\,|\,\theta\in[0,2\pi)\}$.
An unambiguous parametrization of $\P$ is then given 
with the same parameters as above but restricting their domains to be either
$\theta\in[0,\pi)$ and $r\in\R$, or $\theta\in[0,2\pi)$ and $r\ge 0$. 
From now on we use the double covering space $\S\times\R$ instead of $\P$ and, for example,
identify functions $\P\to\C$ with functions $f:\,[0,2\pi)\times\R\to\C$ with the property $f(-\theta,-r)=f(\theta,r)$ for all
$\theta\in[0,2\pi)$ and $r\in\R$. Moreover, without any further mention, we equip $\S\times\R$ with the measure $(2\pi)^{-1}d\theta \,dr$.

Let $\cal S(\P)$ rapidly decreasing smooth functions $\P\to\C$ (obviously in the direction of growing $r$). Let
$\cal S_H(\P)$ consist of functions $\fii\in\cal S(\P)$ such that, for all $k\in\Z_+$, the mapping $\theta\mapsto\int_\R\fii(\theta,r)r^kdr$ is a homogeneous polynomial of degree $k$ with respect to 'variables' $\cos\theta$ and $\sin\theta$.

The {\it Radon transformation} is a continuous $(\star)$ linear mapping
$
\Ra:\,L^1(\R^2)\to L^1(\S\times\R)
$
given by
$$
(\Ra f)(\theta,r):=\int_\R f(r\cos\theta-t\sin\theta,\,r\sin\theta+t\cos\theta)dt
$$
for $d\theta dr$-almost all $(\theta,r)$;
here we consider the above $L^1$-spaces as Banach spaces equipped with the stardard $L^1$-norms.
We have the following Schwartz theorem \cite[theorem 2.4]{He}:
$$
{\Ra}|_{\cal S(\R^2)}:\,\cal S(\R^2)\to\cal S_H(\P)\text{ is bijective.} 
$$

The operator $\Ra$ has an adjoint operator, the so-called {\it backprojection operator}, $\Ra^*$ defined as
$$
(\Ra^*\fii)(x,y):=\int_0^{2\pi}\fii(\theta,\,x\cos\theta+y\sin\theta)\frac{d\theta}{2\pi}
$$
where, e.g., $\fii\in L^1(\S\times\R)$. Indeed,
for suitable classes of functions $f:\,\R^2\to\C$ and $\fii:\,\S\times\R\to\C$
one can easily verify $(\star)$ that
$
\<\Ra f|\fii\>_{L^2(\S\times\R)}
=\<f|\Ra^*\fii\>_{L^2(\R^2)}.
$



Let $f,\,g\in L^1(\R^2)$, and define the {convolution} $f*g\in L^1(\R^2)$ as
$$
(f*g)(x,y):=\iint_{\R^2}f(x-t,y-u)g(t,u)dtdu.
$$
We have the following {\it convolution theorem} $(\star)$:
\begin{equation}\label{co}
[\Ra(f*g)](\theta,r)=\int_\R(\Ra f)(\theta,r-s)(\Ra g)(\theta,s)ds.
\end{equation}

Define the {\it Hilbert transform} $\H$ which is a bounded linear operator on $L^2(\R)$ given by
$$
(\H\psi)(r):=\frac{1}{\pi}\lim_{\epsilon\to0+}\int_{x\in\R\atop|r-x|>\epsilon}\frac{\psi(x)}{r-x}dx
=-\frac{1}{\pi}\lim_{\epsilon\to0+}\int_\epsilon^\infty\frac{\psi(r+y)-\psi(r-y)}{ y}d y
$$
for almost all $r\in\R$ and the above limit exists also in the $L^2$-norm \cite[theorem 8.1.7]{Bu}.\footnote{
Note that some authors multiply $\H$ by the imaginary unit $i$ in the definition of the Hilbert transform to get a self-adjoint operator.}
The Hilbert transform has the following properties:
\begin{enumerate}
\item $\H$ is shift invariant $(\star)$, that is,
$$
(\H\psi_a)(r)=(\H\psi)(r-a),\hspace{1cm}a,\,r\in\R, 
$$
where $\psi\in L^2(\R)$ and $\psi_a(x):=\psi(x-a)$ for all $x\in\R$.
\item $\H$ commutes with derivation \cite[prop.\ 8.3.7]{Bu}, that is,
$$
\H\left(\frac{d\psi}{d x}\right)=\frac{d \H\psi}{d r},
$$
where $\psi\in L^2(\R)$ is absolutely continuous and $d\psi/dx \in L^2(\R)$  (that is, when $\psi$ belong to the domain of the momentum operator $P$).
\end{enumerate}

For any $\fii\in\cal S_H(\P)$ define $\fii_\theta(x):=\fii(\theta,x)$ and
\begin{equation}\label{lambda}
(\Lambda\fii)(\theta,r):=\pi \H\left(\frac{d\fii_\theta}{d x}\right)(r)=
\lim_{\epsilon\to0+}\int_{x\in\R\atop|r-x|>\epsilon}\frac{1}{r-x}\frac{\partial\fii(\theta,x)}{\partial x}dx
=\pi\frac{d }{d r}(\H\fii_\theta)(r).
\end{equation}

We have the {\it Radon inversion theorem} \cite[theorem 3.6]{He}: for all $f\in\cal S(\R^2)$,
\begin{equation}\label{in}
f=\frac{1}{2\pi}\Ra^*\Lambda\Ra f,
\end{equation}
which implies the following {\it Plancherel formula:}
\begin{eqnarray}\nonumber
\iint_{\R^2} \overline{f(x,y)}g(x,y)dxdy&=&\<f|g\>_{L^2(\R^2)}=
\frac{1}{2\pi}\<f|\Ra^*\Lambda\Ra g\>_{L^2(\R^2)}=
\frac{1}{2\pi}\<\Ra f|\Lambda\Ra g\>_{L^2(\S\times\R)} \\
&=&
\frac{1}{4\pi^2}\int_\R\int_0^{2\pi}\overline{(\Ra f)(\theta,r)}(\Lambda\Ra g)(\theta,r)d\theta dr \label{Pl}
\end{eqnarray}
for all $f,\,g\in\cal S(\R^2)$.
\newline

\noindent{\bf Acknowledgments.} The author thanks Pekka Lahti for valuable discussions and for the carefully reading of the manuscript.

\end{document}